\documentclass[%
 reprint,
 amsmath,amssymb,
 aps,
]{revtex4-1}

\usepackage{graphicx}
\usepackage{dcolumn}
\usepackage{bm}

\usepackage{float}

\begin{document}

\preprint{APS/123-QED}

\title{Curvature generation in nematic surfaces}%

\author{Cyrus Mostajeran}
\affiliation{%
Department of Engineering, University of Cambridge, Cambridge CB2 1PZ, United
Kingdom }%

\date{\today}

\begin{abstract}

In recent years there has been a growing interest in the study of shape
formation using modern responsive materials that can be preprogrammed to undergo
spatially inhomogeneous local deformations. In particular, nematic liquid crystalline
solids offer exciting possibilities in this context. Considerable recent
progress has been made in achieving a variety of shape transitions in thin
sheets of nematic solids by engineering isolated points of concentrated
Gaussian curvature using topological defects in the nematic
director field across textured surfaces. In this paper, we consider ways of achieving shape
transitions in thin sheets of nematic glass by generation of non-localised 
Gaussian curvature in the absence of topological defects in the director field.
We show how one can blueprint any desired Gaussian curvature in a thin
nematic sheet by controlling the nematic alignment angle across the surface and
highlight specific patterns which present feasible initial targets for
experimental verification of the theory.
\end{abstract}
 
\maketitle 


\section{Introduction}

Complex shape transitions driven by inhomogeneous local deformation patterns are
ubiquitous in nature and are observed at a variety of length scales, ranging from the cell
walls of plants and bacteria to macrostructures such as plant leaves
\cite{dervaux2008morphogenesis}.
Examples of such local deformations include differential growth in biological structures
and nonuniform mechanical responses of biological tissues to external stimuli
such as humidity \cite{goriely2011morphoelasticity}. Modern responsive materials
can be preprogrammed to undergo spatially inhomogeneous expansions and contractions in 
response to external
stimuli such as heat and light. Examples include thermoresponsive hydrogels
\cite{klein2007shaping} and nematic liquid crystalline solids
\cite{modes2011gaussian}. There has been considerable interest in recent years
in the prospect of preprogramming desired shape transformations that can be remotely activated in such
materials \cite{efrati2009elastic,modes2011blueprinting,kim2012designing}.  

Liquid crystalline solids consist of long, flexible molecular chains that are
sufficiently cross-linked to form a solid. At sufficiently high temperatures,
the molecular directions are randomly distributed and the material is in the
isotropic phase. When the temperature is below some critical value, the
rod-like molecular elements become locally aligned about the director
$\boldsymbol{n}$ and the material is said to be in the nematic phase.
Liquid crystalline solids experience elongations and contractions in response to light, heat, pH, and
other stimuli that change the molecular order. Of particular significance are
nematic glasses \cite{van2007glassy} and elastomers \cite{warner2003liquid}.
Both have spontaneous deformation tensors of the form
\begin{equation}
F=(\lambda-\lambda^{-\nu})\boldsymbol{n}\otimes\boldsymbol{n}+\lambda^{-\nu}\,\mathrm{Id}_3,
\end{equation}
where $\mathrm{Id}_3$ denotes the identity operator on $\mathbb{R}^3$.
That is, a local scaling by $\lambda$ along the director $\boldsymbol{n}$ and a
scaling by $\lambda^{-\nu}$ perpendicular to $\boldsymbol{n}$ is observed upon
exposure to stimulus. The parameter $\nu$ is known as the opto-thermal Poisson
ratio and relates the perpendicular and parallel responses
\cite{modes2011blueprinting}. 
 
In nematic glasses, molecular chain motion is highly limited by the cross-links
and the director is not independently mobile from the elastic matrix as it is in
elastomers. For nematic glasses, we typically have $\lambda\in(0.96,1.04)$ and
$\nu\in(\frac{1}{2},2)$, while for elastomers $\lambda\in(\frac{1}{2},4)$ and
$\nu=\frac{1}{2}$.
In nematic glasses, the director field changes only through convection by
mechanical deformations, which allows for a feasible patterning of the director
field at the initial time of cross-linking and the subsequent guarantee that the
pattern will not be modified by ``soft elasticity'' mediated by director
rotation \cite{warner2003liquid}.
Considerable control of director fields is achievable through a variety of techniques including
the use of electric and magnetic fields, surface anchoring, and holography.
Thus, the matrial response in nematic glasses can be preprogrammed by precisely
setting up a desired director field pattern immediately before the glass is
formed via cross-linking \cite{modes2011blueprinting,van2007glassy}. 

It has come to our attention that a
recent ambitious paper by Aharoni \emph{et al.}
\cite{aharoni2014geometry} addresses the question of shape transformations that
are achievable through the patterning of general smooth director fields on flat
sheets, as was the original motivation for much of this work. Nonetheless, we hope
that this paper will add to their
contribution by highlighting specific patterns which present feasible
initial targets for experimental confirmations of the theory. We also
introduce the notion of orthogonal duality of director fields and consider
its implication for
Gaussian curvature distributions. Nematic patterns on initially curved surfaces
are considered towards the end of the paper.

The value of $\lambda$ and $\nu$ are typically dependent
on the strength of the stimulus. In the following analysis, we assume that the
shape transformation is achieved by a sudden exposure to stimulus of fixed
strength, so that there is a sudden activation of the prescribed local
deformations for fixed values $\lambda$ and $\nu$. For definiteness, we assume
that $\lambda<1$, so that the local deformations consist of a contraction along
the director and an elongation in the perpendicular directions. We use the
convention that Latin indices run over 1, 2, 3, whereas Greek indices take
values 1 and 2.

\section{Non-Euclidean local deformations and nematic surfaces}

A configuration of a body in $\mathbb{R}^3$ can be specified by a smooth injective 
immersion $\boldsymbol{\Phi}:\Omega\rightarrow\mathbb{R}^3$, where $\Omega\subset\mathbb{R}^3$ 
is a domain. The configuration defines curvilinear coordinates $x=(x^i)=(x^1,x^2,x^3)\in\Omega$ 
for material points throughout the body. The map $\boldsymbol{\Phi}:\Omega\rightarrow\mathbb{R}^3$ 
induces a Riemannian metric $g$ on $\Omega$ with 
covariant components
\begin{equation}
g_{ij}(x)=\partial_i\boldsymbol{\Phi}\cdot\partial_j\boldsymbol{\Phi}.
\end{equation}
The Christoffel symbols are defined in terms of the components of the metric tensor by 
\begin{equation}
\Gamma_{ijl}=\frac{1}{2}(\partial_jg_{il}+\partial_ig_{jl}-\partial_lg_{ij})
\end{equation}
and $\Gamma^k_{ij}=g^{kl}\Gamma_{ijl}$, where $(g^{ij})=(g_{ij})^{-1}$. It is
well-known that the components
\begin{equation}
R_{lijk}=\partial_j\Gamma_{ikl}-\partial_k\Gamma_{ijl}+\Gamma^m_{ij}\Gamma_{klm}-\Gamma^m_{ik}\Gamma_{jlm}
\end{equation}
of the Riemann curvature associated with any such induced metric uniformly
vanish at all points in $\Omega$ \cite{ciarlet2005introduction}.

Processes such as differential growth in biological structures can determine 
a reference geometry that is characterised by
a \emph{reference Riemannian metric} $\bar{g}$ that describes the prescribed 
rest distances associated with the underlying 
growth law. Typically, the underlying growth law gives rise to a reference 
metric that is non-Euclidean in the sense that
its Riemann curvature tensor has non-vanishing components
\cite{efrati2009elastic}.

As the body deforms in response to the underlying local deformations,
 it must assume a configuration in three-dimensional 
Euclidean space. Thus, its realised configuration is characterised
by a Riemannian metric $g$ that is induced by a 
configuration $\boldsymbol{\Phi}:\Omega\rightarrow\mathbb{R}^3$. 
The metric $g$ is referred to as the \emph{actual metric}. 
If the underlying local deformations are geometrically incompatible 
with Euclidean space, the prescribed rest distances 
described by $\bar{g}$ cannot be realised everywhere. This generates 
a \emph{residual strain field} throughout the body 
given by
$
\epsilon(x)=\frac{1}{2}\left(g(x)-\bar{g}(x)\right).
$
For a hyperelastic material, the elastic energy is given by a functional
\begin{equation} \label{elastic}
E[\boldsymbol{\Phi}]=\int_{\Omega}W(x,g(x))\,dV,
\end{equation}
where $dV=\sqrt{|\bar{g}|}\,dx_1dx_2dx_3$. At each point $x$, the 
energy density function $W(x,g(x))$ vanishes if and only
if $g(x)=\bar{g}(x)$ \cite{efrati2013metric}.

Consider a standard domain of parametrisation of the form 
$\Omega=\omega\times\left(-\frac{h}{2},\frac{h}{2}\right)$ for a 
thin shell of thickness $h$, where $\omega$ corresponds to the mid-surface and
the third coordinate $x_3\in\left(-\frac{h}{2},\frac{h}{2}\right)$ measures the
distance along the normal to the mid-surface. We can derive a reduced
two-dimensional model formulated in terms of the first and second fundamental forms $a_{\alpha\beta}$, $b_{\alpha\beta}$ of the 
mid-surface $\boldsymbol{\Phi}(\omega)$ by using the Kirchhoff kinematic 
assumption and integrating over the shell thickness $h$. 
In the case of a homogeneous and isotropic elastic material, we obtain 
the reduced energy functional
\begin{align}
\mathcal{E}&=\int_{\omega}\mathcal{W}(x_1,x_2)\sqrt{|\bar{a}|}\,dx_1dx_2
\nonumber \\
&=
\int_{\omega}\left(\mathcal{W}_S(x_1,x_2)+\mathcal{W}_B(x_1,x_2)\right)\sqrt{|\bar{a}|}\,dx_1dx_2
\end{align}
where
\begin{align}
\mathcal{W}_S(x_1,x_2)&=\frac{h}{2}\mathcal{A}^{\alpha\beta\gamma\delta}(a_{\alpha\beta}-\bar{a}_{\alpha\beta})(a_
{\gamma\delta}-\bar{a}_{\gamma\delta}) \\
\mathcal{W}_B(x_1,x_2)&=\frac{h^3}{24}\mathcal{A}^{\alpha\beta\gamma\delta}
(b_{\alpha\beta}-\bar{b}_{\alpha\beta})(b_{\gamma\delta}-\bar{b}_{\gamma\delta})
\end{align}
are the stretching and bending energy contributions, respectively \cite{efrati2013metric}. The
\emph{reference fundamental forms} $\bar{a}_{\alpha\beta}$ and $\bar{b}_{\alpha\beta}$ are related to the 
three-dimensional reference metric $\bar{g}$ via
\begin{equation}
\bar{a}_{\alpha\beta}(x_1,x_2)=\bar{g}_{\alpha\beta}\vert_{x_3=0}, \;\;\;
\bar{b}_{\alpha\beta}(x_1,x_2)=-\frac{1}{2}\partial_3\bar{g}_{\alpha\beta}\vert_{x_3=0}.
\end{equation}
The components of the elasticity tensor $\mathcal{A}^{\alpha\beta\gamma\delta}$
\cite{efrati2013metric} are given by
\begin{align}
\mathcal{A}^{\alpha\beta\gamma\delta}=\frac{Y}{4(1-\nu_{\mathrm{el}}^2)}\big[\nu_{\mathrm{el}}
\bar{a}^{\alpha\beta}
\bar{a}^{\gamma\delta}&
\nonumber
\\
+\;\;\frac{1}{2}(1-\nu_{\mathrm{el}})&(\bar{a}^{\alpha\gamma}
\bar{a}^{\beta\delta}+\bar{a}^{\alpha\delta}\bar{a}^{\beta\gamma})\big],
\end{align}
where the Young modulus $Y$ and Poisson ratio $\nu_{\mathrm{el}}$ are related to
the Lam\'{e} coefficients $\lambda_{\mathrm{el}}$ and $\mu_{\mathrm{el}}$ by
\begin{equation}
\frac{\nu_{\mathrm{el}}}{1-\nu_{\mathrm{el}}}=\frac{\lambda_{\mathrm{el}}}{\lambda_{\mathrm{el}}
+2\mu_{\mathrm{el}}},
\;\;\; \;\;\; Y=2\mu_{\mathrm{el}}\,(1+\nu_{\mathrm{el}}).
\end{equation}
As $h\rightarrow 0$, the equilibrium configuration manifests as an isometric immersion of the reference two-dimensional metric $\bar{a}_{\alpha\beta}$. That is,
\begin{equation}
a_{\alpha\beta}=\bar{a}_{\alpha\beta} \;\;\;\;\;\; \mathrm{as} \;\;\;\;\;\; h\rightarrow 0.
\end{equation}

More precisely, the analysis of the thin sheet limiting behaviour has been
placed on a more rigorous mathematical foundation in recent years using the
concept of $\Gamma$-convergence. In Friesecke
\emph{et al.} \cite{friesecke2002theorem} the nonlinear bending theory of plates
due to Kirchhoff is derived as the $\Gamma$-limit of the classical theory of
3D nonlinear elasticity, under the assumption that the
classical 3D elastic energy per unit thickness $h$ scales like $h^2$. The
non-Euclidean version of this result is derived in \cite{lewicka2011scaling}
under the same scaling law applied to the energy functional (\ref{elastic}),
yielding a natural non-Euclidean generalization of the Kirchhoff model with 
a corresponding 2D bending  energy functional. Necessary and sufficient
conditions for the existence of a $W^{2,2}$ isometric immersion of a given 2D
metric into $\mathbb{R}^3$ are also established. In particular, it is shown that if finite
bending energy isometric immersions of the metric exist, then the minimizers of the 3D
elastic energy (\ref{elastic}) converge in the vanishing thickness limit to an
isometric immersion of the mid-surface metric, which is a global minimizer of the
bending energy over all isometric immersions \cite{lewicka2011scaling}.

Consider a director field $\boldsymbol{n}:\Omega\rightarrow\mathbb{R}^3$ given
by
$\boldsymbol{n}=\boldsymbol{n}(x_1,x_2,x_3)=n_i(x_1,x_2,x_3)\hat{\boldsymbol{e}}^i$,
where $x_i$ are the coordinates with respect to an orthonormal coordinate frame
$\{\boldsymbol{e}^i\}$ on $\Omega$ and $\{\hat{\boldsymbol{e}}^i\}$ denotes the
standard orthonormal basis of Euclidean space. 
The spontaneous deformation tensor is given by
\begin{equation}
\bar{F}_{ij}=\left(\lambda-\lambda^{-\nu}\right)n_in_j+\lambda^{-\nu}\delta_{ij}.
\end{equation}
The reference metric determined by the nematic director field is 
\begin{align}
\bar{g}_{ij} &= (\bar{F}^T)_{ik}(\bar{F})_{kj} \nonumber \\
&= (\lambda-\lambda^{-\nu})^2 n_{k}n_{k}n_{i}n_{j} \nonumber \\
&+\;
2\lambda^{-\nu}(\lambda-\lambda^{-\nu})n_{i}n_{j}
+\lambda^{-2\nu}\delta_{ki}\delta_{kj}.
\end{align}
Since $n_k n_k=1$, we obtain 
\begin{equation} 
\bar{g}_{ij} = (\lambda^2-\lambda^{-2\nu})n_{i}n_{j}+\lambda^{-2\nu}\delta_{ij}.
\end{equation}

Here we are interested in surface director field patterns on initially flat thin
sheets, so we consider a plate of thickness $h$ with a standard domain of parametrisation
of the form $\Omega=\omega\times\left(-\frac{h}{2},\frac{h}{2}\right)$ and director fields of the form
\begin{equation}  \label{eq:director1}
\boldsymbol{n}(x)=n_1(x_1,x_2)\hat{\boldsymbol{e}}^1+n_2(x_1,x_2)\hat{\boldsymbol{e}}^2,
\end{equation} 
where $\omega$ denotes the mid-plate.
That is, we assume that the same surface director field is repeated at each level 
across the plate thickness. The reference metric now has components
\begin{equation}
\bar{g}_{\alpha\beta}=\left(\lambda^2-\lambda^{-2\nu}\right)n_{\alpha}
n_{\beta}+\lambda^{-2\nu}\delta_{\alpha\beta},
\end{equation}
and $\bar{g}_{33}=\lambda^{-2\nu}$, $\bar{g}_{\alpha 3}=0$.
The reference fundamental forms in the reduced two-dimensional model are 
\begin{equation}
\bar{a}_{\alpha\beta}(x_1,x_2)=\bar{g}_{\alpha\beta}(x_1,x_2,0), \;\;\;\;\;
\bar{b}_{\alpha\beta}(x_1,x_2)=0.
\end{equation}
Thus, in the thin sheet limit the components $a_{\alpha\beta}$ of the
first fundamental form of the mid-surface are given by
\begin{equation}
a_{\alpha\beta}=(\lambda^2-\lambda^{-2\nu})n_{\alpha}n_{\beta}+\lambda^{-2\nu}\delta_{\alpha\beta}.
\end{equation}
Since the $2$D in-plane director field is modelled as a normalised vector field,
it can be specified by a single \emph{angle scalar field} $\psi=\psi(x_1,x_2)$ such that
\begin{equation}
n_1=\cos\psi(x_1,x_2), \;\;\;\;\;
n_2=\sin\psi(x_1,x_2).
\end{equation}

Recall that the \emph{Gaussian curvature} $K$ at a point $p$ on a surface is defined as 
the product of the two principal curvatures at $p$. In terms of the fundamental forms 
of the surface, it can be expressed as
$K=\det(b_{\alpha\beta})/\det(a_{\alpha\beta})$.
According to the \emph{Theorem Egregium} of Gauss, the Gaussian curvature is a characteristic 
of the intrinsic geometry of a surface. In particular, it is uniquely determined by the 
first fundamental form $a_{\alpha\beta}$ according to the equation
\begin{align}
K=-\frac{1}{a_{11}}\Big(\partial_1\Gamma^2_{12}&-\partial_{2}\Gamma^2_{11}+\Gamma^1_{12}\Gamma^2_{11}\nonumber
\\
&-\Gamma^1_{11}\Gamma^2_{12}+\Gamma^2_{12}\Gamma^2_{12}-\Gamma^2_{11}\Gamma^2_{22}\Big).
\end{align}
A direct calculation of the Gaussian curvature associated with the nematic metric yields the expression
\begin{align}
K = \frac{1}{2}&\left(\lambda^{2\nu}-\lambda^{-2}\right)\left[
\left(\partial^2_2 \psi -\partial^2_1\psi -4 \partial_1 \psi \partial_2 \psi\right)\sin (2 \psi)\right. \nonumber \\
&+ \;\;
\left.2\left(\partial_1\partial_2\psi+(\partial_2\psi)^2-(\partial_1\psi)^2\right)\cos (2 \psi)\right],
\end{align}
in terms of the angle scalar field $\psi$. 

Define the \emph{orthogonal dual} to a director field 
to be the director field obtained by rotating all the nematic directors by
$\pi/2$.
That is, the dual director field is characterised by the angle field $\tilde{\psi}=\psi+\pi/2$. Upon 
taking the 
orthogonal dual
$\psi \rightarrow \psi+\pi/2$, we find that
$\partial_{\alpha}\psi\rightarrow \partial_{\alpha}\psi$,
$\sin(2\psi)\rightarrow\sin(2\psi+\pi)=-\sin2\psi$ and
$\cos(2\psi)\rightarrow\cos(2\psi+\pi)=-\cos2\psi$, 
so that 
\begin{equation} {\label{dual}}
K\rightarrow -K \;\;\; \mathrm{as} \;\;\; \psi \rightarrow \psi+\pi/2.
\end{equation}
That is, the orthogonal dual of any director field has precisely the opposite Gaussian curvature
at every point.  

\section{Shifted director patterns}

Suppose that the components of the director field depend only on one of the coordinates, so 
that $\boldsymbol{n}=\boldsymbol{n}(x_2)$, say. Such a director field can be specified as
\begin{equation} \label{eq:1angle}
\boldsymbol{n}=\cos\psi(x_2)\,\hat{\boldsymbol{e}}^1+\sin\psi(x_2)\,\hat{\boldsymbol{e}}^2.
\end{equation}
The 
Gaussian curvature of the associated metric is given by
\begin{equation}
K=-\frac{1}{2}\left(\lambda^{-2}-\lambda^{2\nu}\right)\left(\psi''\sin
2\psi+2\psi'^2\cos 2\psi\right).
\end{equation}
For constant $K\in\mathbb{R}$, this yields a second order ordinary differential equation
in $\psi$:
\begin{equation}
\frac{d^2}{d x_2^2}\cos2\psi=4\,C(K),
\end{equation} 
where
$C(K)={K}/(\lambda^{-2}-\lambda^{2\nu})$.
This ODE is solved by
\begin{equation}
\psi(x_2)=\pm\frac{1}{2}\cos^{-1}\Big(c_1+c_2\,x_2+2\,C(K)\,x_2^2\Big),
\end{equation}
where $c_1$, $c_2$ are constants of integration. The corresponding 
director field generates constant Gaussian curvature $K$ 
wherever it is well-defined. Fig.~\ref{fig:1plots} shows examples of director
fields that generate constant Gaussian curvature of various types. 
\begin{figure*}[t!]
\centering
\includegraphics[width=0.95\linewidth]{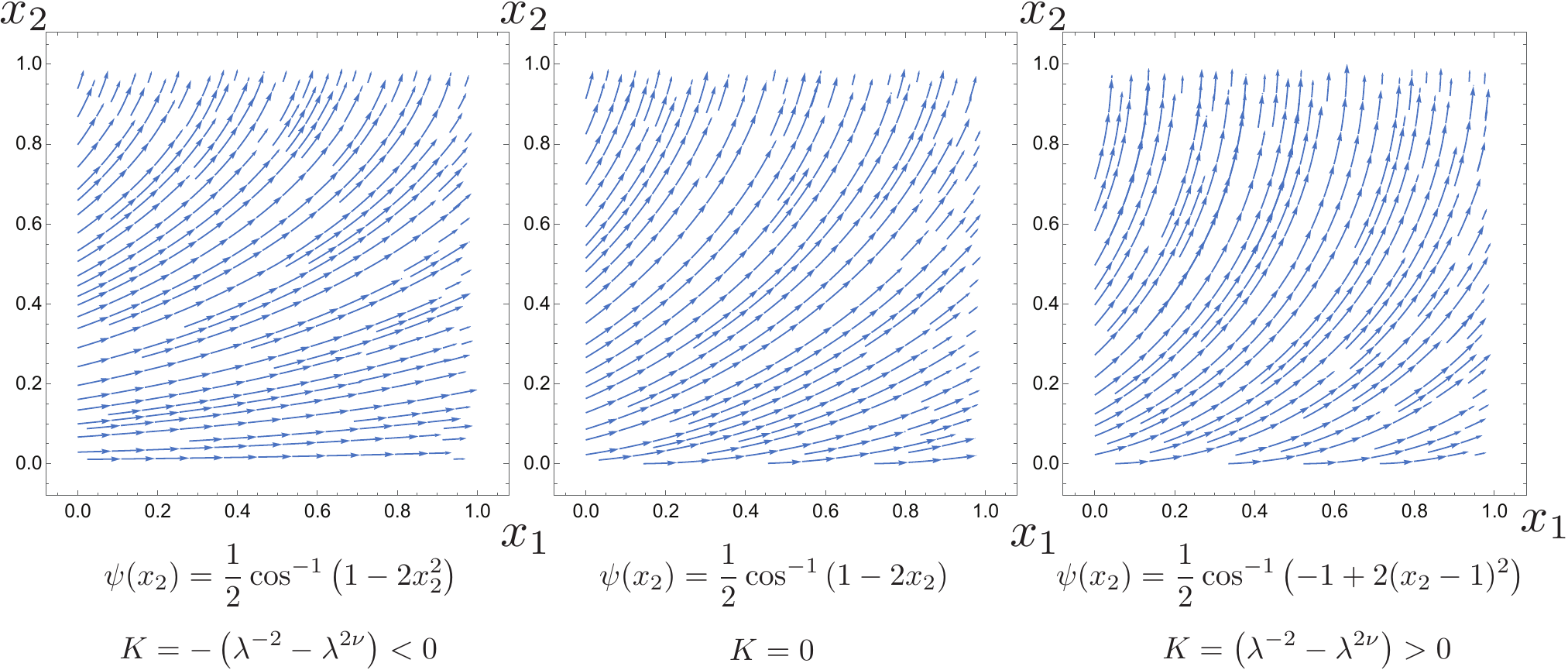}
  \caption{\label{fig:1plots} Director fields of the form
 $\boldsymbol{n}=\cos\psi(x_2)\,\hat{\boldsymbol{e}}^1+\sin\psi(x_2)\,\hat{\boldsymbol{e}}^2$
 on the unit square for different choices of the function $\psi=\psi(x_2)$. The
  resulting Gaussian curvature $K$ upon stimulation is also indicated.}
  \label{fig:1plots}
\end{figure*}

Director fields of the form~(\ref{eq:1angle}) can be
generated by shifting a fixed curve along the $x_1$-axis. We now develop a more geometric
description for director fields that can be generated by uniform translation of 
a curve $\boldsymbol{\gamma}=\boldsymbol{\gamma}(t)=\gamma_1(t)\hat{\boldsymbol{e}}^1+\gamma_2(t)\hat{\boldsymbol{e}}^2$ 
along a fixed direction specified by a unit vector $\boldsymbol{u}=u_1\hat{\boldsymbol{e}}^1+u_2\hat{\boldsymbol{e}}^2$. 
For such a director field, it is natural to change coordinates from the Cartesian $(x_1,x_2)$ to $(t,r)$ where $t$ is 
the parameter along the curve $\boldsymbol{\gamma}$ and $r$ is the parameter in the direction of translation. That is, 
we have
\begin{equation}
x_1(t,r):=\gamma_1(t)+ru_1, \;\;\;\;\;\; x_2(t,r):=\gamma_2(t)+ru_2.
\end{equation}
The director field $\boldsymbol{n}=n_1\hat{\boldsymbol{e}}^1+n_2\hat{\boldsymbol{e}}^2$ at each point $(t,r)$ is given by
\begin{equation}
n_1(t)=\frac{\gamma_1'}{\sqrt{\gamma_1'^2+\gamma_2'^2}}, \;\;\;\;\;\;\; n_2(t)=\frac{\gamma_2'}{\sqrt{\gamma_1'^2+\gamma_2'^2}}.
\end{equation}
Note that the independence of the director field components from $r$ follows by construction. The metric components 
$
\mathbf{A}=[a_{\alpha\beta}]=\left[(\lambda^2-\lambda^{-2\nu})n_{\alpha}n_{\beta}+\lambda^{-2\nu}\delta_{\alpha\beta}\right]
$
in Cartesian coordinates transform according to $\mathbf{A}\rightarrow \mathbf{J}^T\mathbf{AJ}$, where $\mathbf{J}$ 
is the Jacobian matrix
\begin{equation}
\mathbf{J} = \left(
\begin{array}{cc}
 \partial_t x_1 & \partial_{r}x_1 \\
 \partial_t x_2 & \partial_{r} x_2\\
\end{array}
\right).
\end{equation}
That is,
\begin{align}
\left(
\begin{array}{cc}
 a_{tt} & a_{tr} \\
 a_{rt} & a_{rr}\\
\end{array}
\right) = 
\left(
\begin{array}{cc}
 \gamma'_1(t) & u_1 \\
 \gamma'_2(t) & u_2 \\
\end{array}
\right)^T 
\left(
\begin{array}{cc}
 a_{11} & a_{12} \\
 a_{21} & a_{22}\\
\end{array}
\right)
\left(
\begin{array}{cc}
 \gamma'_1(t) & u_1 \\
 \gamma'_2(t) & u_2 \\
\end{array}
\right) \nonumber \\
\end{align}

Now for a given curve $\boldsymbol{\gamma}$ and specified direction
$\boldsymbol{u}$, we can compute the Gaussian curvature as a scalar field
$K=K(t)$ using the equation
\begin{align}
K=-\frac{1}{a_{tt}}\Big(\partial_t\Gamma^{r}_{tr}&-\partial_{r}\Gamma^{r}_{tt}
+\Gamma^t_{tr}\Gamma^{r}_{tt}
\nonumber
\\
&-\Gamma^t_{tt}\Gamma^{r}_{tr}+\Gamma^{r}_{tr}\Gamma^{r}_{tr}-\Gamma^{r}_{tt}\Gamma^{r}_{rr}\Big).
\end{align}
Here we note a remarkable result with an elegant geometric interpretation. Take
a circular arc 
$\boldsymbol{\gamma}(t)=R\left(\cos t, \sin t\right)$,
where 
$R=\frac{1}{\sqrt{K}}\left(\lambda^{-2}-\lambda^{2\nu}\right)^{1/2}$,
and $K>0$ is a constant. Now let $\boldsymbol{u}=(\cos\alpha,\sin\alpha)$, 
where $\alpha$ is some fixed angle. A direct computation of the Gaussian curvature 
associated with the director field generated by translating $\boldsymbol{\gamma}$ 
along $\boldsymbol{u}$, yields $K$ exactly. It is assumed that the
circular arc and direction of translation are chosen such that the director
field does not self-intersect.
  
\begin{figure}[h]  
\centering
\includegraphics[width=1\linewidth]{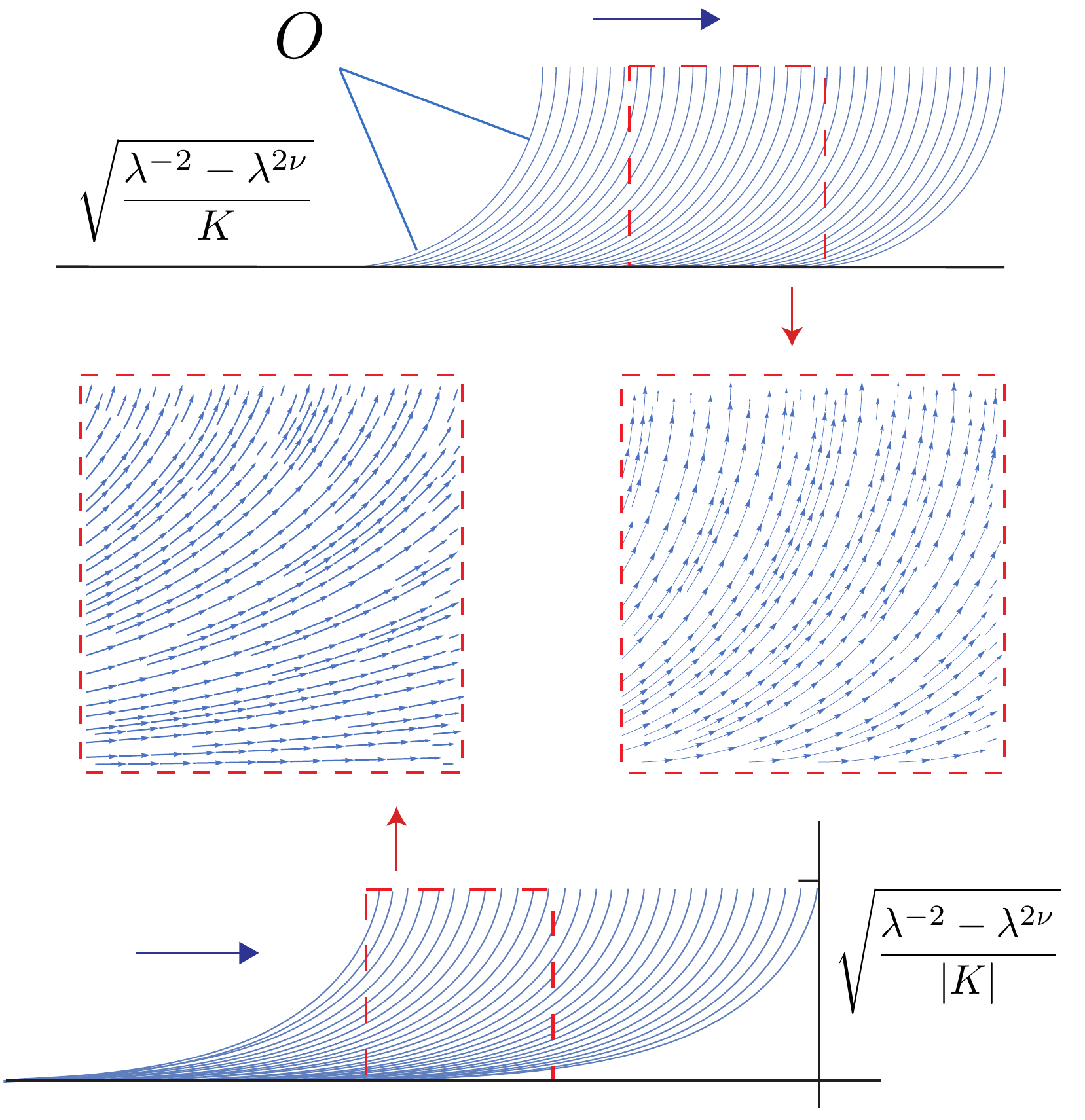}
  \caption{\label{fig:TractShift01} The nematic pattern obtained by shifting a
  circular arc along a fixed direction generates constant positive Gaussian
  curvature $K>0$. The nematic pattern obtained by shifting a tractrix along its
  axis generates constant negative Gaussian curvature $K<0$.}
  \label{fig:ShiftTractrix}
\end{figure}

Recall that the \emph{tractrix} in the
$x_1x_2$-plane whose axis coincides with the $x_1$-axis is the planar curve
$\boldsymbol{\gamma}$ passing through a point $(0,a)$ with the property that the
length of the segment of the tangent line from any point on the curve to the
$x_1$-axis is constant and equal to $a>0$.
One parametric representation of the tractrix is provided by
\begin{equation}  \label{eq:Tractrix}
\gamma_1(t)=a\,(\tanh t-t),  \;\;\;\;\;\;
\gamma_2(t)=a\,\mathrm{sech}\, t.
\end{equation}
We now consider the director pattern that is generated by translating a tractrix
\emph{along its axis}. That is, we take $\boldsymbol{\gamma}$ to be as in
equation (\ref{eq:Tractrix}) and choose the direction of translation to be
$\boldsymbol{u}=(1,0)$ as shown in Fig.~\ref{fig:TractShift01}.
A direct computation of the Gaussian curvature of such a pattern generated by a tractrix 
with parameter
$a=\frac{1}{\sqrt{|K|}}\left(\lambda^{-2}-\lambda^{2\nu}\right)^{1/2}$ 
with $K<0$, yields $K$ exactly.
  
\section{Orthogonal Duality}

\begin{figure}[H]  
\centering
\includegraphics[width=0.9\linewidth]{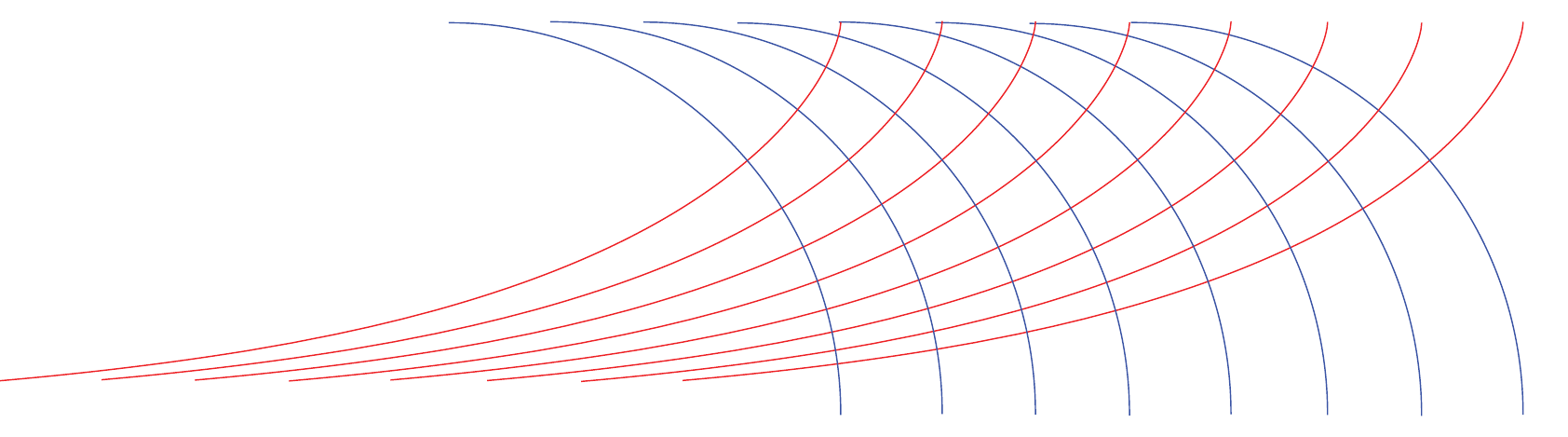}
  \caption{\label{fig:orthogonal2}Orthogonal duality of director fields. The
  director fields that were discovered to generate constant Gaussian curvatures
  of opposite sign are seen to be equivalent to orthogonally dual patterns. Note
  that every tractrix intersects every circle at right angles.}
  \label{fig:circleTesselation} 
\end{figure}

The result that the director fields in Fig.~\ref{fig:TractShift01} generate
opposite Gaussian curvature may seem surprising on first inspection,
since the patterns look somewhat similar. However, in light of the orthogonal
duality result (\ref{dual}), we find that the patterns make perfect sense. In
particular, the positive curvature pattern is equivalent to the pattern obtained
by rotating the sheet by $90^{\circ}$. One can see that this equivalent pattern
is exactly the orthogonal dual of the negative curvature pattern of
Fig.~\ref{fig:TractShift01}, as illustrated in Fig
~\ref{fig:orthogonal2}. 

This observation is consistent with physical intuition since the
mechanical response of nematic solids to stimuli such as heat is a contraction
along the director and a simultaneous expansion in the orthogonal directions.
Thus, one might expect that the intrinsic curvature properties of
a patterned sheet will be qualitatively reversed when the pattern is replaced by
its orthogonal dual, even though the scaling factors $\lambda$ and $\lambda^{-\nu}$ 
are not exact reciprocals when $\nu\neq 1$.
Indeed, we have shown that this reversal is exact in the sense that the 
Gaussian curvature distribution becomes precisely
the negative of the dual pattern.

It is also interesting to connect these ideas to the mechanical
response of a patterned nematic sheet that is being cooled instead of heated. On
cooling an initially flat nematic sheet, the rod-like molecules expand along
the director and contract in the orthogonal directions. So once again one may expect
the activated surface to exhibit the opposite of the curvature properties that
manifest on heating. Quantifying this opposite response however is somewhat
delicate, since the scaling coefficients are typically temperature dependent and
cooling a flat sheet by a particular temperature will not necessarily produce
the exact opposite of the response observed by heating the flat sheet by
the same temperature. 

\section{Willmore Functional}

It is well known that for a surface in $\mathbb{R}^3$, the components $a_{\alpha\beta}$ and $b_{\alpha\beta}$ 
of the first and second fundamental forms satisfy the Gauss-Codazzi-Mainardi equations
\begin{align}
\partial_\beta\Gamma_{\alpha\sigma\tau} -
\partial_\sigma\Gamma_{\alpha\beta\tau} +
\Gamma^{\mu}_{\alpha\beta}\Gamma_{\sigma\tau\mu} - 
\Gamma^{\mu}_{\alpha\sigma}&\Gamma_{\beta\tau\mu} \nonumber \\
&=
b_{\alpha\sigma}b_{\beta\tau} - b_{\alpha\beta}b_{\sigma\tau}, \nonumber \\
\partial_{\beta}b_{\alpha\sigma} - \partial_{\sigma}b_{\alpha\beta} +
\Gamma^{\mu}_{\alpha\sigma}b_{\beta\mu} -
\Gamma^{\mu}_{\alpha\beta}b_{\sigma\mu}& = 0,
\end{align}
where
$\Gamma_{\alpha\beta\tau} =
\frac{1}{2}(\partial_{\beta}a_{\alpha\tau}+\partial_{\alpha}a_{\beta\tau}
-\partial_{\tau}a_{\alpha\beta})$ and
$\Gamma^{\sigma}_{\alpha\beta}= a^{\sigma\tau}\Gamma_{\alpha\beta\tau}$.
Furthermore, any pair $(a,b)$ consisting of a
symmetric and positive definite matrix field $(a_{\alpha\beta})$ and a symmetric
matrix field $(b_{\alpha\beta})$ that satisfy the Gauss-Codazzi-Mainardi
equations determines a unique surface up to a rigid transformation 
in $\mathbb{R}^3$ \cite{ciarlet2005introduction}.

To determine the equilibrium configuration of the mid-surface of an initially flat nematic sheet upon stimulation, 
we need to know the components $b_{\alpha\beta}$ of the second fundamental form that minimise the bending energy
\begin{equation}
\mathcal{E}[b]=\frac{1}{3}\int_{\omega}\mathcal{A}^{\alpha\beta\gamma\delta}b_{\alpha\beta}b_{\gamma\delta}\,dS
\end{equation}
subject to the Gauss-Codazzi-Mainardi constraints defined by the metric $(a_{\alpha\beta})$ generated by the 
director field. 
This functional can be rewritten in terms of the mean and Gaussian curvatures of
the mid-surface as 
\begin{equation}
\mathcal{E}=\frac{1}{3}\int_{\omega}\left(\frac{4H^2}{1-\nu_{\mathrm{el}}}-2K\right)dS.
\end{equation}
Since $a_{\alpha\beta}=\bar{a}_{\alpha\beta}$, and the Gaussian curvature $K$ is an isometric invariant, 
the problem reduces to minimising the \emph{Willmore functional} 
\begin{equation}
I_W=\int_{\omega}H^2\,dS
\end{equation}
among isometric immersions of the metric
\cite{efrati2011hyperbolic,willmore2012introduction,lewicka2011scaling}.
\newline 

\begin{figure}[h]  
\centering 
\includegraphics[width=0.8\linewidth]{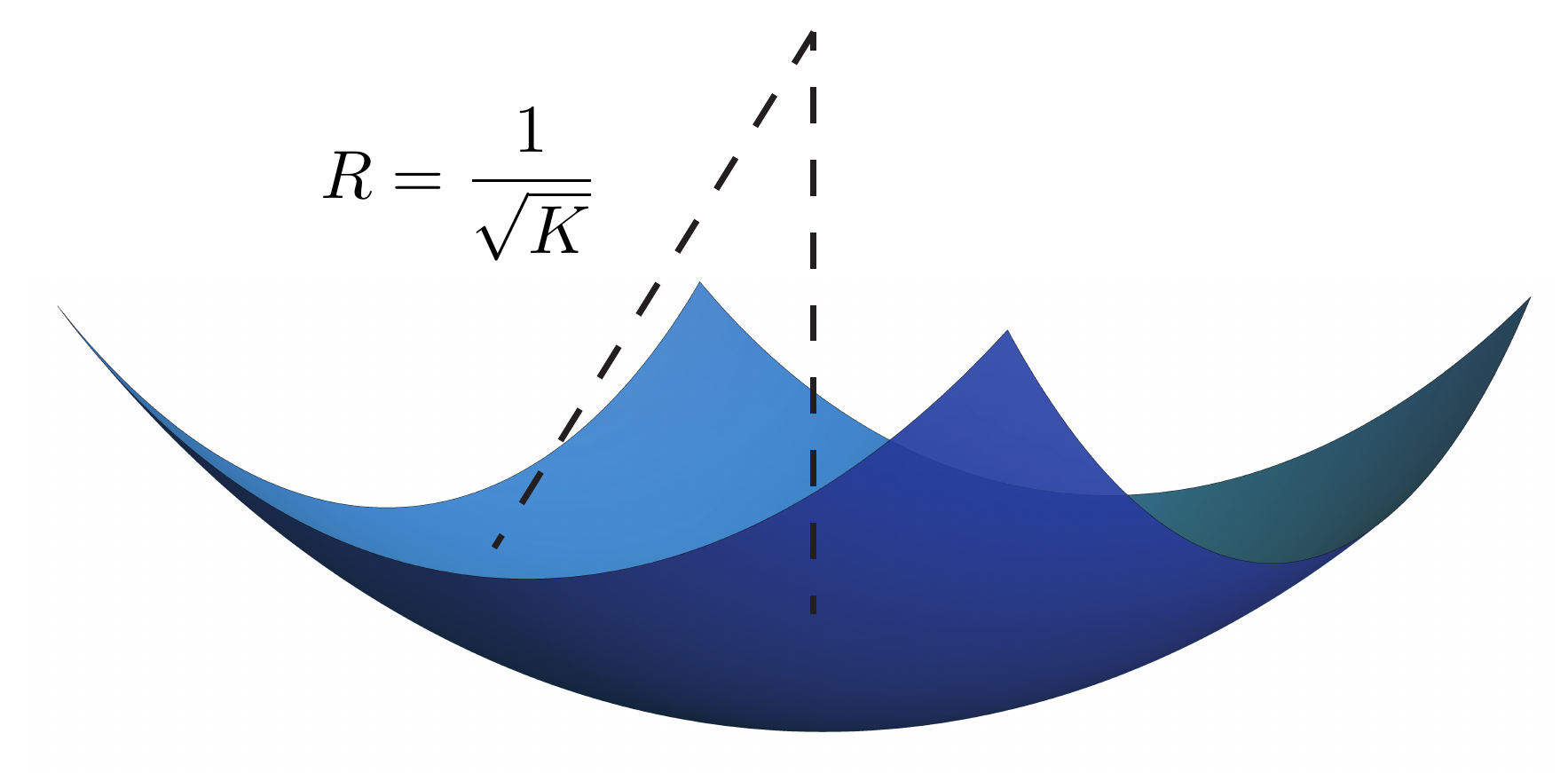}
  \caption{\label{fig:sphereformation}An initially flat nematic sheet forming
  part of a sphere upon stimulation.} 
  \label{fig:sphere}
\end{figure}

The Willmore functional can be written in terms of the principal curvatures $\kappa_1$, $\kappa_2$ as
\begin{align}
I_W&=\frac{1}{4}\int_{\omega}(\kappa_1+\kappa_2)^2dS  \nonumber \\
&=\frac{1}{4}\int_{\omega}(\kappa_1-\kappa_2)^2dS + \int_{\omega}K\,dS,
\end{align}
where $K=\kappa_1\kappa_2$ is the Gaussian curvature. We note that if the
condition $\kappa_1=\kappa_2$ throughout $\omega$ is consistent with the metric,
then the equilibrium configuration corresponding to minimal bending energy is
determined by imposing this condition.
The sphere is the only compact surface in $\mathbb{R}^3$ whose principal curvatures are equal
everywhere \cite{willmore2012introduction}. Since surfaces of the same
constant Gaussian curvature $K$ are locally isometric by Minding's theorem
\cite{minding}, a flat nematic sheet whose director field encodes constant
positive curvature $K$ will form part of a sphere of radius $R=1/\sqrt{K}$ upon stimulation, assuming that the sheet is small
enough to exclude the possibility of self-intersection
as in Fig.~\ref{fig:sphereformation}.
\newline

In the case of a thin sheet encoded with constant negative curvature $K<0$, the
study of the equilibrium shapes demands a considerably more
delicate analysis. We follow Gemmer \emph{et al.} \cite{gemmer2011shape} and
consider equilibrium configurations of hyperbolic elastic disks of radius $R$ that
have already undergone local deformations. The equilibrium configurations in the
thin sheet limit correspond to minimisers of the Willmore functional or,
equivalently, to minimisers of
\begin{equation} \label{energy}
\tilde{I}_W=\int_{\omega}\left(\kappa_1^2+\kappa_2^2\right)dS,
\end{equation}
over isometric immersions of the metric.
In geodesic polar coordinates $(\rho,\theta)$, where the metric takes the form
$ds^2=d\rho^2+\left|K\right|^{-1}\sinh^2\left(\sqrt{\left|K\right|}\rho\right)d\theta^2$,
Eq. (\ref{energy}) can be written as
\begin{equation}
\tilde{I}_W=\int_{0}^{2\pi}\int_0^R\;\frac{\sinh\left(\sqrt{\left|K\right|}\rho\right)}{\sqrt{\left|K\right|}}
\left(\kappa_1^2+\kappa_2^2\right)d\rho \,d\theta.
\end{equation}

It is well-known that at every point of a hyperbolic surface, there exists a
pair of asymptotic curves along which the normal curvature vanishes
\cite{han2006isometric}.
Furthermore, the isometric immersions of the hyperbolic metric of constant
Gaussian curvature $K$ are in correspondence with solutions of the sine-Gordon
equation:
\begin{equation}
\frac{\partial^2\phi}{\partial u \, \partial
v}=\left|K\right|\sin\phi(u,v),
\end{equation}
where $u$, $v$ are coordinates for a local parametrisation by asymptotic curves
and $\phi$ denotes the angle between the asymptotic curves. The angle $\phi$ is
related to the principal curvatures by
\begin{equation}
\kappa_1^2=\left|K\right|\tan^2\frac{\phi}{2}, \;\;\;\;\;
\kappa_2^2=\left|K\right|\cot^2\frac{\phi}{2},
\end{equation}
so that 
\begin{align} \label{functional}
\tilde{I}_W =
\int_{0}^{2\pi}\int_0^R\;\frac{\sinh\left(\sqrt{\left|K\right|}\rho\right)}{\sqrt{\left|K\right|}}
\left(\tan^2\frac{\phi}{2}+\cot^2\frac{\phi}{2}\right)d\rho\,d\theta.
\end{align}

In \cite{gemmer2011shape}, the minimisation of Eq.
(\ref{functional}) over all \emph{smooth} solutions to the sine-Gordon equation
is analysed numerically to conclude that the principal curvatures of smooth
isometric immersions of a hyperbolic disk of radius $R$ satisfy
\begin{equation}
\max\left(|\kappa_1|,|\kappa_2|\right)\geq\frac{|K|}{64}e^{2\sqrt{\left|K\right|}R}.
\end{equation}
Moreover, it is shown that this lower bound is attained by surfaces that are
geodesic disks lying on hyperboloids of revolution of constant Gaussian
curvature $K$. 

Hyperbolic surfaces whose two asymptotic curves are straight lines that
intersect at an angle $\theta$ form a one-parameter family of surfaces
$\mathcal{A}_{\theta}$ known as Amsler surfaces \cite{amsler}. Each such surface
is uniquely determined by the angle $\theta$ between the asymptotic lines. In
\cite{gemmer2011shape}, the odd periodic extension of a subset of the Amsler
surface $\mathcal{A}_{\pi/n}$ bounded between the asymptotic lines is taken to
generate a periodic profile with $n$ waves. The resulting $n$-periodic shapes
are referred to as periodic Amsler surfaces $A_n$, which are not smooth as they
have discontinuities in their second derivatives at the lines of inflection.

Numerical investigations of the bending energy associated with geodesic disc
cutouts of $A_n$ have revealed that for large values of
$\epsilon=\sqrt{\left|K\right|}R$, particular $n$-wave periodic Amsler surfaces
are energetically more favorable than the smooth saddle shapes that correspond to
discs lying on hyperboloids of revolution. Furthermore, it has been shown
that there exists a sequence of critical radii $R_n\sim\ln(n)$ beyond which the
principal curvatures and hence Willmore functional of $A_n$ diverge. If the
radius of the hyperbolic disc is sufficiently large for periodic Amsler
surfaces to be energetically more favorable than the hyperboloid solutions, the
particular $n$-wave surface that is expected to form is predicted by identifying
the smallest critical radius $R_n$ greater than $R$. Thus, as the radius of the hyperbolic
disc increases one observes that $n$-periodic shapes with an increasing number
of waves become more energetically favorable \cite{gemmer2011shape,
gemmer2013shape}. 

It is interesting to contrast the case of shape selection of sheets encoded with
constant positive curvature with that of sheets patterned with a hyperbolic metric.
In the constant positive curvature case, the analysis suggests that
subsets of spheres are the most energetically favorable isometric
immersions provided that the dimensions of the initial sheet are small enough to
rule out self-intersection of the deformed surface. That is, the nature of the 
solution is not influenced
by the shape and size of the initially flat sheet. In the hyperbolic
case, however, we notice that the extrinsic geometry of the deformed surface
is crucially dependent on the particular shape and size of the initially flat
sheet. In particular, surfaces with dramatically different extrinsic geometries
may form by simply increasing the size of the sample patterned with a particular
intrinsic geometry.

\section{Blueprinting nematic sheets of prescribed Gaussian curvature}

\begin{figure}[h] 
\centering
\includegraphics[width=0.95\linewidth]{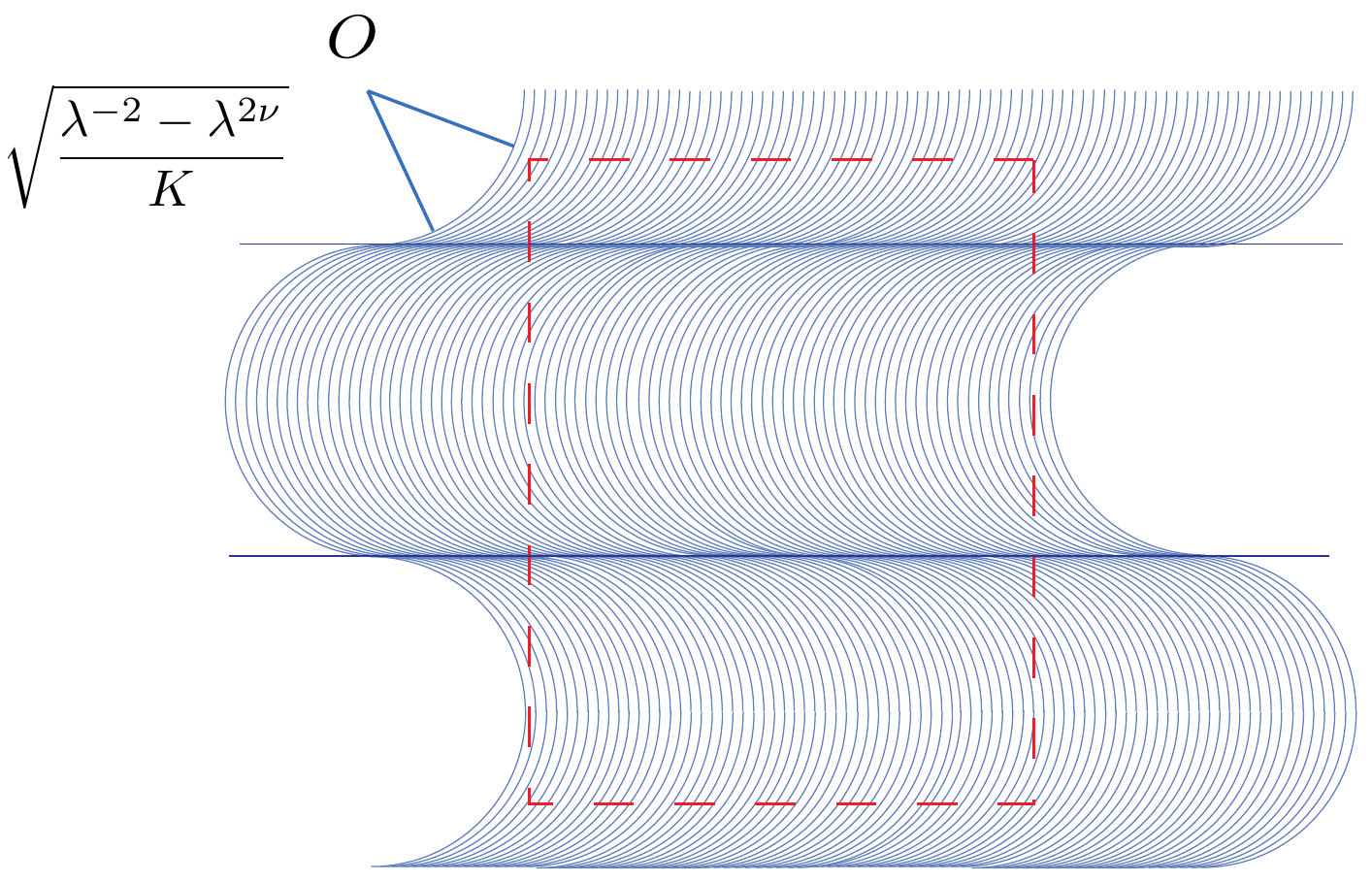} 
  \caption{\label{fig:extend1}Blueprinting prescribed positive Gaussian
  curvature $K>0$ on a rectangular sheet of arbitrary dimensions whose
  edges are indicated by the dashed lines.
  The Gaussian curvature is well-defined everywhere except along the solid
  horizontal lines where the distinct arcs join.}
  \label{fig:circleTesselation} 
\end{figure}

One can use the shifted circular arc patterns of Fig.~\ref{fig:TractShift01} to
blueprint a metric of constant positive Gaussian curvature that is realised upon
stimulation of a thin nematic sheet. Since the magnitude of the Gaussian curvature is determined
by the radius of the shifted circular arcs, we need to use a shifted pattern
consisting of circular arcs that are $C^1$-smoothly joined in order to extend
the pattern to cover larger domains, as shown in Fig.~\ref{fig:extend1}. 
\newline
 
\begin{figure}[h] 
\centering
\includegraphics[width=0.95\linewidth]{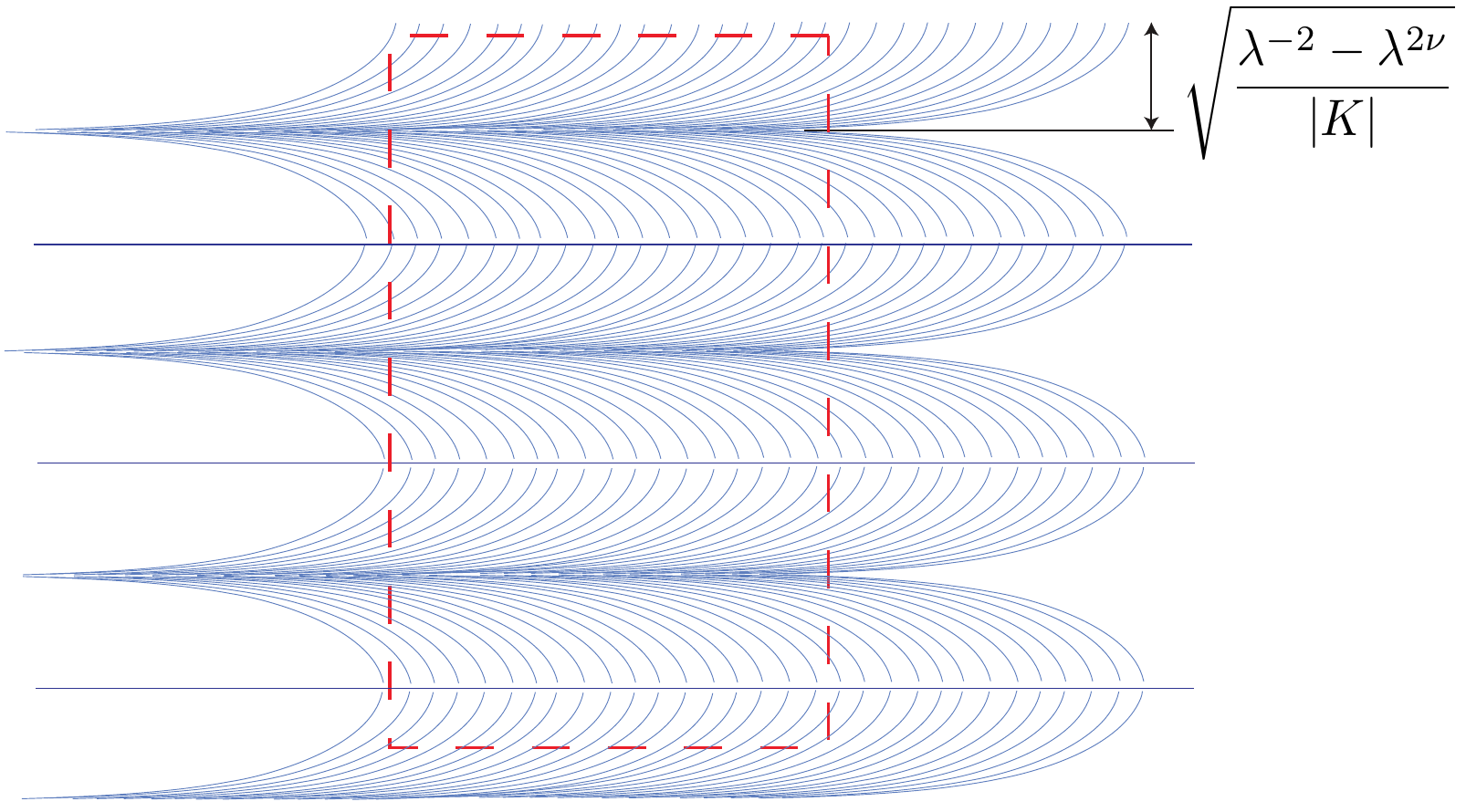}
  \caption{\label{fig:extend2}Blueprinting prescribed negative Gaussian
  curvature $K<0$ on a rectangular sheet of arbitrary dimensions whose
  edges are indicated by the dashed lines.
  The Gaussian curvature is well-defined everywhere except along the solid
  horizontal lines.}
  \label{fig:circleTesselation} 
\end{figure}

The pattern on the larger domain encodes positive Gaussian curvature at
all points except along the lines where the distinct circular arcs join.
At these points, the Gaussian curvature is not defined and sharp creases may
develop upon stimulation. Similarly, one can extend the shifted tractrix patterns 
of Fig.~\ref{fig:TractShift01} to encode constant negative curvature on
extended domains as shown in Fig.~\ref{fig:extend2}. As in the positive
curvature case, the Gaussian curvature is not well-defined along the
solid horizontal lines indicated in Fig.~\ref{fig:extend2} as the metric
components are not continuously differentiable at these points. 

In the case of the circular arc patterns, for instance, note that the unit
tangents to the arcs give the director field whose first derivative at each
point coincides with the vector pointing towards the centre of the circular arc
at that point and so clearly exhibits a jump discontinuity across the solid
horizontal line. The jump discontinuities of the extended hyperbolic pattern
follow by orthogonal duality.

\section{Nematic patterns on curved surfaces}

Often we are interested in shape transitions of surfaces that are initially
curved. In such cases the surface may have a non-flat first fundamental form at
the outset, which would then change due to the activation of some local
deformation pattern. In the context of thin nematic shells, we would start with
a shell whose mid-surface is given as an embedding in three dimensions. This
embedding determines the first fundamental form as a non-trivial $(0,2)$ metric
tensor, which would then change to a new reference metric
$\bar{a}_{\alpha\beta}$ that depends on the director field pattern
on the surface.

Here we highlight an interesting result on the effect of two sets of director
fields on the Gaussian curvature of surfaces of revolution. Consider a surface
of revolution whose standard parametrisation is given by
\begin{equation}
\boldsymbol{r}(u,v)=\left(f(v)\cos u, f(v)\sin u, g(v)\right),
\end{equation}
where the pair of real-valued functions $(f,g)$ specifies the profile curve of
the surface. The associated metric is given by
\begin{align}
ds^2&=\boldsymbol{r}_u\cdot\boldsymbol{r}_u
du^2+2\boldsymbol{r}_u\cdot\boldsymbol{r}_v du \, dv
+\boldsymbol{r}_v\cdot\boldsymbol{r}_v dv^2 \nonumber \\
&= f^2du^2+\left(f'^2+g'^2\right)dv^2.
\end{align}
A direct calculation of the Gaussian curvature yields
\begin{equation}
K=\frac{f'g'g''-f''g'^2}{f(f'^2+g'^2)^2}.
\end{equation}
Now suppose that the surface is patterned such that the directors are aligned
with the $v=\mathrm{constant}$ lines. That is, the director field lines are
circles centred on the axis of revolution. The metric upon stimulation becomes
\begin{equation}
ds^2=f^2 \; \lambda^2 du^2 + \left(f'^2+g'^2\right)\;\lambda^{-2\nu}dv^2,
\end{equation}
and the Gaussian curvature transforms as
\begin{equation}
K=\lambda^{2\nu}\frac{f'g'g''-f''g'^2}{f(f'^2+g'^2)^2}.
\end{equation}
That is, for any surface of revolution patterned by director fields aligned with
the horizontal circles centred on the axis of revolution, the
Gaussian curvature transforms according to the elegant formula $K\rightarrow
\lambda^{2\nu}K$. This result was established in \cite{modes2012responsive} for
the special case of a sphere patterned by azimuthal director fields. Here we see
that the result is valid for all surfaces of revolution, including surfaces of
negative and non-constant Gaussian curvature. 

For the orthogonal dual of the considered pattern on any given surface of
revolution, the metric upon stimulation is given by
\begin{equation}
ds^2=f^2 \; \lambda^{-2\nu} du^2 + \left(f'^2+g'^2\right)\;\lambda^2 dv^2,
\end{equation}
and a direct calculation of the Gaussian curvature yields
\begin{equation}
K=\lambda^{-2}\frac{f'g'g''-f''g'^2}{f(f'^2+g'^2)^2}.
\end{equation}
That is, for any surface of revolution patterned by director fields aligned with
the profile curves of the surface (i.e. lines of constant $u$), the Gaussian
curvature transforms according to $K\rightarrow \lambda^{-2}K$. Note the
remarkable fact that this transformation is independent of the opto-thermal
Poisson ratio $\nu$. In the case of a sphere, such patterns correspond to the
director fields being aligned with the lines of longitude.

In the special case of a sphere of radius $R$, these results show that upon the
stimulation of an azimuthal pattern the Gaussian curvature decreases to
$\lambda^{2\nu}/R^2$, whereas for the orthogonal dual pattern the Gaussian
curvature increases to $\lambda^{-2}/R^2$. We now show how these two patterns and the
associated Gaussian curvature responses can be viewed as the extreme cases
of the family of loxodromic spiral patterns on a sphere, as shown in
Fig.~\ref{fig:new3}. 
 
\begin{figure*}[t!]
\centering
\includegraphics[width=1\linewidth]{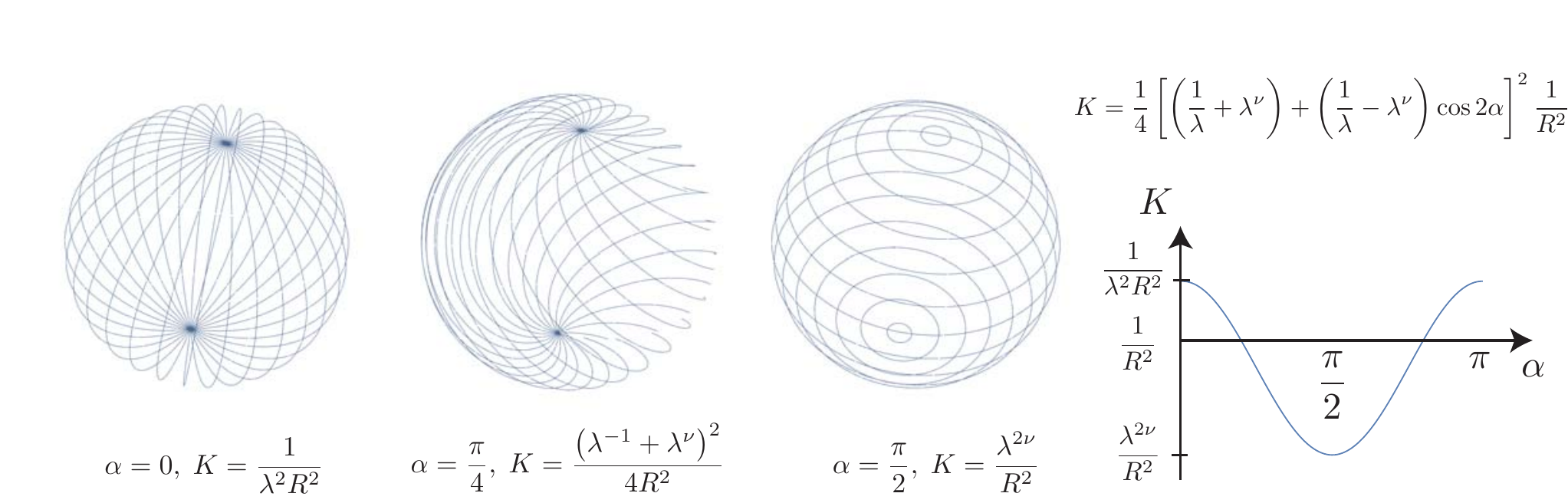}
  \caption{\label{fig:new3} Loxodromic spiral director fields of the form
  $\boldsymbol{n}=
  \cos\alpha\,\hat{\boldsymbol{e}}_{\theta}+\sin\alpha\,\hat{\boldsymbol{e}}_{\phi}$
  on a sphere of radius $R$ for the cases $\alpha=0, \, \pi/4,\, \pi/2$ and
  the Gaussian curvature generated upon stimulation. Note that $\alpha$ is the
  angle that the directors make with the meridian lines.}
  \label{fig:new3}
\end{figure*}

Recall that 
a loxodrome on a sphere is the curve that cuts a meridian at a constant angle
$\alpha$. If $\alpha$ is not a right angle, then the loxodromes form spirals on
the sphere. We assume that a sphere of radius $R$ is patterned with a director
field whose integral curves consist of loxodromes of angle $\alpha$, so that
$\boldsymbol{n}=
\cos\alpha\,\hat{\boldsymbol{e}}_{\theta}+\sin\alpha\,\hat{\boldsymbol{e}}_{\phi}$,
where $\theta$ and $\phi$ are the polar and azimuthal angles, respectively, and
$\hat{\boldsymbol{e}}_{\theta}$, $\hat{\boldsymbol{e}}_{\phi}$ are the
corresponding unit tangent vectors in $\mathbb{R}^3$. Upon stimulation, the point $(\theta,\phi)$
transforms according to $(\theta,\phi)^T \rightarrow F (\theta,\phi)^T$, where
$F$ is the matrix representation of the spontaneous deformation tensor with respect to the
orthonormal basis $\hat{\boldsymbol{e}}_{\theta}$,
$\hat{\boldsymbol{e}}_{\phi}$.
The round spherical metric $ds^2= R^2\left(d\theta^2+\sin^2\theta \,d\phi^2\right)$
transforms as
\begin{equation}
ds^2\rightarrow \begin{pmatrix}d\theta & d\phi\end{pmatrix} F^Ta_0F
\begin{pmatrix}d\theta
\\
d\phi\end{pmatrix} = \begin{pmatrix}d\theta & d\phi\end{pmatrix} a
\begin{pmatrix}d\theta
\\ 
d\phi\end{pmatrix},
\end{equation}
where 
\begin{equation}
a_0=\begin{pmatrix}R^2 & 0 \\  0 & R^2 \sin^2\theta \end{pmatrix}.
\end{equation}
Directly computing the Gaussian curvature using the new metric $a=F^Ta_0F$,
we obtain the formula
\begin{equation}
K=\frac{1}{4}\left[\left(\frac{1}{\lambda}+\lambda^{\nu}\right)+
\left(\frac{1}{\lambda}-\lambda^{\nu}\right)\cos2\alpha\right]^2\frac{1}{R^2}.
\end{equation}

We notice that the resulting Gaussian curvature is a constant
between the values $\lambda^{2\nu}/R^2$ and $\lambda^{-2}/R^2$ depending
on the angle $\alpha$. Note that although the Gaussian curvature upon
stimulation is constant and positive in each case, the resulting surface is
generally not a sphere. For example, in the case of azimuthal patterns
($\alpha=\pi/2$) a spindle or ``thorny sphere" of the corresponding curvature is
expected to form \cite{modes2012responsive}.
\newline

\section{Comments}     

It is often the case that key physical characteristics of
materials, such as adhesive and optical properties, are determined by surface
structure. The engineering of switchable surfaces consisting of active materials such as
nematic glasses offers the possibility of controllably and reversibly
changing the surface geometry of thin structures for use 
in a variety of potential devices. Possible 
applications may include microfluidic mixers and pumps, adjustable optical
lenses and switchable textured surfaces for use in tablet computers. There
are two key mechanisms for generating shape transformations in thin sheets of
patterned nematic glasses. One method is by means of defects in the
director field pattern and the other is by using a smooth in-plane director
field that generates Gaussian curvature. 

In this paper, we have clearly highlighted smooth director fields that generate
constant Gaussian curvature of any desired value, with the hope that such patterns offer
feasible initial targets for experimentalists to confirm the predictions of the
theory and test its limitations. We have also identified an important symmetry
in the Gaussian curvature associated with a given nematic pattern on an
initially flat sheet, namely that the orthogonal dual of a given pattern
generates the exact opposite Gaussian curvature at every point. 
This observation has an important practical
implication for experimentalists, who can produce a sheet whose Gaussian
curvature distribution is the exact opposite of a given patterned sheet by
simply adding $\pi/2$ radians to the nematic alignment angle associated with the
given sheet. If the viability of producing
switchable nematic surfaces using such patterns is confirmed, one may begin to
effectively combine more elaborate smooth director fields and topological
defects to devise a variety of powerful shaping mechanisms for use in a wide
range of potential applications.
 
\begin{acknowledgments}
The author is most grateful to Professor Mark Warner of Cavendish Laboratory for
his guidance and support. Insightful comments and suggestions from the anonymous
reviewers are appreciatively acknowledged as having improved the quality of this
paper. The author is supported by the Engineering and Physical
Sciences Research Council of the United Kingdom. The computations
were performed in \emph{Mathematica 10.0} using the author's code. 
\end{acknowledgments}

\nocite{*}

\bibliography{document}

\end{document}